\documentstyle[preprint,aps]{revtex}
\begin{document}

\draft

\title{Universal Behavior of Quantum Impurity Scattering in 
Tomonaga-Luttinger Liquid}

\author{Yu-Liang Liu}
\address{Department of Physics, Hong Kong University of Science and 
Technology, Clear Water Bay, Kowloon, Hong Kong, People's Republic of China}

\maketitle

\begin{abstract}

Using bosonization and path integral methods, we study general low temperature
behavior of non-magnetic and magnetic impurity scattering in 
Tomonaga-Luttinger liquid, and calculate electron Green function for a
general backward scattering potential. We demonstrate that electron 
density of state near the impurity site is suppressed by the backward 
scattering, but it mainly remains invariant as far away from the impurity, and
at zero temperature the electrons are completely reflected on the impurity
site, the system breaks into two subsystems but right- and left-moving
electron fields have a twisted boundary condition. We also show that a testing
charge can only be partially screened by conduction electrons, and in strong
interaction region the impurity susceptibility has a $1/T$-type low
temperature behavior.

\end{abstract}
\vspace{1cm}


\newpage

\section{Introduction}

Recently, considerable efforts have been directed towards the study of the 
Fermi edge singularity\cite{1,2,3,4,5,6,7}, the Kondo effect\cite{8,9,10,11} 
and the transport properties of one-dimensional(1D) 
Tomonaga-Luttinger(TL)-liquids\cite{12,13,14,15,16,17,18,19,20}.
The common property of a magnetic and a non-magnetic impurity scattering in
TL-liquid is that there exists backward scattering of electrons on impurity,
which drastically influences the low energy behavior of the system. The
rigorous treatment of the backward scattering is a hard work because in low
energy limit the backward scattering potential is renormalized to infinity,
usual perturbation expansion method cannot be directly used.
Just as shown in Ref.\cite{12}, for a repulsive
interacting electronic system, in the low energy limit, the conduction
electrons are completely reflected on the impurity site due to the backward
scattering, the system breaks into two subsystems. Under this consideration, 
the authors in Ref.\cite{21,22} studied the low energy behavior of a
TL-liquid with an open boundary condition $\psi_{R\sigma}(x)=\pm
\psi_{L\sigma}(-x)$, and obtained some results which are consistent with
renormalization group calculation\cite{12}. 
The backward scattering also drastically changes the Fermi edge
singularity\cite{3,4,5,6,7}, which contributes a finite quantity to the
exponent of the X-ray absorption line shape function, although between
Refs.\cite{3,4,5,6} and Ref.\cite{7} there exists some controversy about the
contribution size to this exponent by the backward scattering term.
Although the topics has been extensively studied, 
there are still some hot debating theoretically about that whether is
the electron's density of state enhanced or suppressed near and far away from 
the impurity site? whether
does the system break into two subsystems at the impurity site with the
boundary condition $\psi_{R\sigma}(x)=\pm\psi_{L\sigma}(-x)$ in low energy
limit? and so on. Due to the backward scattering term is relevant,  
we need a method rigorously to treat it. In this paper, using an
elementary method which can rigorously treat the backward scattering term, we
try to clarify these debating points. 

In sections $I\!I$ and $I\!I\!I$, we consider non-magnetic and magnetic 
impurity scattering, respectively. Using an unitary transformation, we
can eliminate the backward scattering term (for magnetic impurity, the
$J^{z}_{2k_{F}}$-term), and incorporate its influence on the system into
electron interaction terms. In section $I\!V$, combining bosonization and
path integral methods, we exactly calculate the Green functions of electrons
$\bar{\psi}_{R(L)\sigma}(x)$ and fermions $\bar{\psi}_{1(2)\sigma}(x)$ for
generally backward scattering, and show that the density of state of electrons
near the impurity is suppressed by the backward scattering, while the electron
density of state far away from the impurity remains intact. In section $V$,
we show that at zero temperature electrons are completely reflected on the 
impurity, the system breaks into two subsystems but the electron fields
have a twisted boundary condition. We calculate the exponent of Fermi-edge
singularity function of X-ray absorption in section $V\!I$. In sections
$V\!I\!I$ and $V\!I\!I\!I$, we show that a testing charge is only partially
screened by the conduction electrons, and study the low temperature
behavior of impurity susceptibility, respectively. We give our conclusion
and some discussion in section $I\!X$.

\section{A non-magnetic impurity scattering}

We consider the following impurity scattering in a general
one-dimensional interacting electron system
\begin{eqnarray}
H_{T} &=& \displaystyle{ H+H_{im}} \nonumber \\
H &=& \displaystyle{-i\hbar v_{F}\sum_{\sigma}\int dx
[\psi^{\dagger}_{R\sigma}(x)\partial_{x}\psi_{R\sigma}(x)
-\psi^{\dagger}_{L\sigma}(x)\partial_{x}\psi_{L\sigma}(x)]} \label{1} \\
&+& \displaystyle{V_{1}\sum_{\sigma}\int dx 
\rho_{R\sigma}(x)\rho_{L\sigma}(x)+V_{2}\sum_{\sigma}\int dx 
\rho_{R\sigma}(x)\rho_{L-\sigma}(x)} \nonumber \\
H_{im} &=& \displaystyle{\sum_{\sigma}
V_{2k_{F}}[\psi^{\dagger}_{R\sigma}(0)\psi_{L\sigma}(0)+
\psi^{\dagger}_{L\sigma}(0)\psi_{R\sigma}(0)]}
\nonumber\end{eqnarray}
where $\psi_{R\sigma}(x)$ and $\psi^{\dagger}_{R\sigma}(x)$ 
are the annihilation and creation 
field operators of the electrons with spin $\sigma$
that propagate to the right with wave vectors $\sim+k_{F}$,
$\psi_{L\sigma}(x)$
and $\psi^{\dagger}_{L\sigma}(x)$ are the annihilation and creation field 
operators of left propagating electrons with spin
$\sigma$ and wave vectors $\sim -k_{F}$; 
$\rho_{R(L)\sigma}(x)=\psi^{\dagger}_{R(L)\sigma}(x)\psi_{R(L)\sigma}(x)$ 
are the electron 
density operators; the spectrum of the electrons is linearized near the 
Fermi points and $v_{F}$ is the Fermi velocity.  
$V_{2k_{F}}=V(k=2k_{F})$ is the backward scattering potential of an
impurity at $x=0$ on the conduction electrons. For simplicity
we have omitted the forward scattering potential because it is trivial in our
following transformations.
In the bosonic representation of the electron fields\cite{24,25,26} 
$\psi_{R(L)\sigma}(x)=(\frac{D}{2\pi\hbar
v_{F}})^{1/2}\exp\{-i\Phi_{R(L)\sigma}(x)\}$, 
where $D$ is the band width of the conduction electrons 
(for simplicity we have neglected the factors $\exp\{\pm ik_{F}x\}$), 
the Hamiltonian $H$ can be written as a diagonal form
\begin{eqnarray}
H &=& \displaystyle{\frac{\hbar v_{c}}{4\pi}\int dx\{
\frac{1}{g_{c}}[\partial_{x}\tilde{\Phi}_{-c}(x)]^{2}+g_{c}
[\partial_{x}\tilde{\Phi}_{+c}(x)]^{2}\}} \nonumber \\
&+& \displaystyle{\frac{\hbar v_{s}}{4\pi}\int dx\{
\frac{1}{g_{s}}[\partial_{x}\tilde{\Phi}_{-s}(x)]^{2}+g_{s}
[\partial_{x}\tilde{\Phi}_{+s}(x)]^{2}\}}
\label{2}\end{eqnarray}
where, $v_{c}=v_{F}(1-\gamma_{c}^{2})^{1/2}$,
$v_{s}=v_{F}(1-\gamma_{s}^{2})^{1/2}$,
$\gamma_{c}=\frac{V_{1}+V_{2}}{2\pi\hbar v_{F}}$,
$\gamma_{s}=\frac{V_{1}-V_{2}}{2\pi\hbar v_{F}}$,
$g_{c}=(\frac{1-\gamma_{c}}{1+\gamma_{c}})^{1/2}$,
$g_{s}=(\frac{1-\gamma_{s}}{1+\gamma_{s}})^{1/2}$,
$\tilde{\Phi}_{\pm c}(x)=\frac{1}{2}[\tilde{\Phi}_{\pm\uparrow}(x)
+\tilde{\Phi}_{\pm\downarrow}(x)]$,
$\tilde{\Phi}_{\pm s}(x)=\frac{1}{2}[\tilde{\Phi}_{\pm\uparrow}(x)
-\tilde{\Phi}_{\pm\downarrow}(x)]$,
$\tilde{\Phi}_{\pm\sigma}(x)=\Phi_{R\sigma}(x)\pm
\Phi_{L\sigma}(x)$. The impurity scattering term can be written
\begin{equation}
H_{im}=\frac{2D}{\pi\hbar v_{F}}V_{2k_{F}}
\cos[\tilde{\Phi}_{-s}(0)]\cos[\tilde{\Phi}_{-c}(0)]
\label{3}\end{equation}
which has a conformal dimension
$(g_{c}+g_{s})/2$. For the repulsive electron-electron interactions with
$V_{1}\geq V_{2}$, the dimensionless coupling strength parameters $g_{c}$
and $g_{s}$ are less than one, the backward scattering term therefore is
relevant, in the low energy limit, 
$V_{2k_{F}}$ is renormalized to be infinity\cite{12}, usual perturbation 
expansion of $V_{2k_{F}}$ is invalid.

In order effectively to study the physical property of the system, 
we define a set of new fermion field operators to diagonalize the Hamiltonian
$H_{im}$
\begin{equation}
\psi_{1\sigma}(x)=\frac{1}{\sqrt{2}}(\psi_{R\sigma}(x)+
\psi_{L\sigma}(-x)), \;\;\;
\psi_{2\sigma}(x)=\frac{1}{\sqrt{2}}(\psi_{R\sigma}(x)-\psi_{L\sigma}(-x))
\label{4}\end{equation}
It is easy to check that the operators $\psi_{1(2)\sigma}(x)$ 
satisfy the standard anticommutation relations. 
In terms of these new fermion fields  
$\psi_{1(2)\sigma}(x)$, the Hamiltonian (1) can be 
written as
\begin{eqnarray}
H_{1} &=& \displaystyle{-i\hbar v_{F}\sum_{\sigma}\int dx
[\psi^{\dagger}_{1\sigma}(x)\partial_{x}\psi_{1\sigma}(x)
+\psi^{\dagger}_{2\sigma}(x)\partial_{x}\psi_{2\sigma}(x)]} \nonumber \\
&+& \displaystyle{\frac{V_{1}}{4}\sum_{\sigma}
\int dx [\rho_{1\sigma}(x)+\rho_{2\sigma}(x)]
[\rho_{1\sigma}(-x)+\rho_{2\sigma}(-x)]} \nonumber \\
&+& \displaystyle{\frac{V_{2}}{4}\sum_{\sigma}
\int dx 
[\rho_{1\sigma}(x)+\rho_{2\sigma}(x)]
[\rho_{1-\sigma}(-x)+\rho_{2-\sigma}(-x)]} \nonumber \\
H_{2} &=& \displaystyle{
-\frac{V_{1}}{4}\sum_{\sigma}\int dx
[\psi^{\dagger}_{1\sigma}(x)\psi_{2\sigma}(x)+h.c.]
[\psi^{\dagger}_{1\sigma}(-x)\psi_{2\sigma}(-x)+h.c.]} \label{5} \\
&-& \displaystyle{
\frac{V_{2}}{4}\sum_{\sigma}\int dx
[\psi^{\dagger}_{1\sigma}(x)\psi_{2\sigma}(x)+h.c.]
[\psi^{\dagger}_{1-\sigma}(-x)\psi_{2-\sigma}(-x)+h.c.]} \nonumber \\
H_{im} &=& \displaystyle{\sum_{\sigma}V_{2k_{F}}
[\rho_{1\sigma}(0)-\rho_{2\sigma}(0)]}
\nonumber \end{eqnarray}
where, $H=H_{1}+H_{2}$. The backward scattering term $H_{im}$ becomes a
very simple form. In order to study low energy behavior of the system where
renormalized backward scattering potential $V^{R}_{2k_{F}}$ goes to infinity
in low energy limit, we rewrite the backward scattering term as the following
form
\begin{equation}
H_{im}=\sum_{\sigma}\hbar v_{F}\delta[\rho_{1\sigma}(0)-\rho_{2\sigma}(0)]
\end{equation}
where $\delta=\arctan(V_{2k_{F}}/(\hbar v_{F}))$ is a phase 
shift induced by the backward scattering potential $V_{2k_{F}}$. 
It is reduced to the $H_{im}$ in (\ref{5}) as $V_{2k_{F}}\rightarrow 0$.
Taking this replacement, we can 
study the property of the system for any value of $V_{2k_{F}}$. 
The Hilbert space of the fields
$\psi_{R(L)\sigma}(x)$ is different from that of the fields
$\psi_{1(2)\sigma}(x)$ in which the backward scattering term becomes the
usual potential scattering in quantum mechanics. However, the transformation
(\ref{4}) is valid for any $V_{2k_{F}}$, we can take the phase shift
$\delta$ as a renormalized quantity varying from zero to $\pm\pi/2$.  
This replacement can be justified by the following facts: a). It is usually
used in treatment of Kondo problem and is proved to be correct. b). In the
interaction-free case, using the Bethe Ansatz it can be shown that the 
impurity scattering potential dependence of ground-state energy is in the form
of phase shift $\delta$. c). In the phase shift description, at the strong
coupling critical points $\delta^{c}=\pm\pi/2$ (corresponding to infinity
backward scattering potential) we can easily calculate the Green function
and density of state of electrons, and show (see below) that they are 
completely consistent with previous calculations\cite{12,21}.
The bosonic representation of the fermion fields 
$\psi_{1(2)\sigma}$ can be written as
$\psi_{1(2)\sigma}(x)=(\frac{D}{2\pi\hbar v_{F}})^{1/2}
\exp\{-i\Phi_{1(2)\sigma}(x)\}$,
where 
$\rho_{1(2)\sigma}(x)=\psi^{\dagger}_{1(2)\sigma}(x)\psi_{1(2)\sigma}(x)$ 
are the density operators, and are related to the
boson field $\Phi_{1(2)\sigma}(x)$ through
$\partial_{x}\Phi_{1(2)\sigma}(x)=2\pi\rho_{1(2)\sigma}(x)$. 

Performing the unitary transformation
\begin{equation}
U=\exp\{i\sum_{\sigma}\frac{\delta}{2\pi}
[\Phi_{1\sigma}(0)-\Phi_{2\sigma}(0)]\}
\label{7}\end{equation}  
we can have the relation
\begin{equation}
U^{\dagger}(H_{1}+H_{2}+H_{im})U=H_{1}+
U^{\dagger}H_{2}U 
\end{equation}
where the unitary transformation of the Hamiltonian $H_{2}$ can be written as
\begin{eqnarray}
\bar{H}_{2} &=& U^{\dagger}H_{2}U \nonumber\\
&=& -\displaystyle{\frac{1}{4}\sum_{\sigma}\int dx\{
V_{1}[e^{-i\delta\;sgn(x)}\psi^{\dagger}_{1\sigma}(x)\psi_{2\sigma}(x)
+h.c.]} \nonumber\\
&\cdot& \displaystyle{
[e^{i\delta\;sgn(x)}\psi^{\dagger}_{1\sigma}(-x)\psi_{2\sigma}(-x)
+h.c.]} \nonumber\\
&+& \displaystyle{V_{2}
[e^{-i\delta\;sgn(x)}\psi^{\dagger}_{1\sigma}(x)\psi_{2\sigma}(x)
+h.c.]}  \\
&\cdot& \displaystyle{
[e^{i\delta\;sgn(x)}\psi^{\dagger}_{1-\sigma}(-x)\psi_{2-\sigma}(-x)
+h.c.]} \nonumber
\end{eqnarray}
Taking the gauge transformations
\begin{equation}
\psi_{1\sigma}(x)=\bar{\psi}_{1\sigma}(x)e^{i\theta_{1}}, \;\;\;
\psi_{2\sigma}(x)=\bar{\psi}_{2\sigma}(x)e^{i\theta_{2}}, \;\;\;
\theta_{1}-\theta_{2}=\pm\delta
\label{9}\end{equation}
which makes $\bar{H}_{2} $ be more compact, and leaves 
$H_{1}$ intact, the Hamiltonian $\bar{H}_{2}=\tilde{H}_{2}+\tilde{H}^{'}_{2}$ 
can be rewritten as 
\begin{eqnarray}
\tilde{H}_{2} &=& 
\displaystyle{-\frac{V_{1}\cos(2\delta)}{4}\sum_{\sigma}\int dx
[\bar{\psi}^{\dagger}_{1\sigma}(x)\bar{\psi}_{2\sigma}(x)+h.c.]
[\bar{\psi}^{\dagger}_{1\sigma}(-x)\bar{\psi}_{2\sigma}(-x)+h.c.]} 
\nonumber \\
&-& \displaystyle{\frac{V_{2}\cos(2\delta)}{4}\sum_{\sigma}\int dx
[\bar{\psi}^{\dagger}_{1\sigma}(x)\bar{\psi}_{2\sigma}(x)+h.c.]
[\bar{\psi}^{\dagger}_{1-\sigma}(-x)\bar{\psi}_{2-\sigma}(-x)+h.c.]} 
\label{10} \\
\tilde{H}^{'}_{2} &=& \displaystyle{i\frac{V_{1}\sin(2\delta)}{2}
\sum_{\sigma}\int^{\infty}_{0}dx
[\bar{\psi}^{\dagger}_{1\sigma}(x)\bar{\psi}_{2\sigma}(x)-h.c.]
[\bar{\psi}^{\dagger}_{1\sigma}(-x)\bar{\psi}_{2\sigma}(-x)+h.c.]} 
\nonumber \\
&+& \displaystyle{i\frac{V_{2}\sin(2\delta)}{2}
\sum_{\sigma}\int^{\infty}_{0}dx
[\bar{\psi}^{\dagger}_{1\sigma}(x)\bar{\psi}_{2\sigma}(x)-h.c.]
[\bar{\psi}^{\dagger}_{1-\sigma}(-x)\bar{\psi}_{2-\sigma}(-x)+h.c.]} 
\nonumber\end{eqnarray}
To further simplify the Hamiltonian $\bar{H}_{2}$, we can
re-define the left- and right-moving electron fields
\begin{equation}
\bar{\psi}_{R\sigma}(x)= \frac{1}{\sqrt{2}}
[\bar{\psi}_{1\sigma}(x)+\bar{\psi}_{2\sigma}(x)], 
\;\;\;
\bar{\psi}_{L\sigma}(-x)= \frac{1}{\sqrt{2}}
[\bar{\psi}_{1\sigma}(x)-\bar{\psi}_{2\sigma}(x)] 
\label{11}\end{equation}
The Hamiltonian $\tilde{H}^{'}_{2}$ becomes
\begin{eqnarray}
\tilde{H}^{'}_{2} &=& 
\displaystyle{\pm i\frac{V_{1}\sin(2\delta)}{2}
\sum_{\sigma}\int^{\infty}_{0}dx
[\bar{\psi}^{\dagger}_{L\sigma}(-x)\bar{\psi}_{R\sigma}(x)-h.c.]
[\bar{\rho}_{R\sigma}(-x)-\bar{\rho}_{L\sigma}(x)]} 
\nonumber \\
&\pm& \displaystyle{i\frac{V_{2}\sin(2\delta)}{2}
\sum_{\sigma}\int^{\infty}_{0}dx
[\bar{\psi}^{\dagger}_{L\sigma}(-x)\bar{\psi}_{R\sigma}(x)-h.c.]
[\bar{\rho}_{R-\sigma}(-x)-\bar{\rho}_{L-\sigma}(x)]} 
\nonumber\end{eqnarray}
where $\bar{\rho}_{R(L)\sigma}(x)=\bar{\psi}^{\dagger}_{R(L)\sigma}(x)
\bar{\psi}_{R(L)\sigma}(x)$ are the density operators.
The Hamiltonian $\tilde{H}^{'}_{2}$ has a conformal dimension $\Delta >1$
because the field $\bar{\rho}_{R\sigma}(x)-\bar{\rho}_{L\sigma}(-x)$ has
the conformal dimension one and the fields $\bar{\psi}^{\dagger}_{L\sigma}
(-x)\bar{\psi}_{R\sigma}(x)$ and $\bar{\psi}^{\dagger}_{R\sigma}(x)
\bar{\psi}_{L\sigma}(-x)$ have the conformal dimension $\Delta'>1$ which can
be seen from Eq.(\ref{22}),
we can neglect it as a first order approximation. It is zero at the strong
coupling critical points $\delta^{c}=\pm\pi/2$, 
$\tilde{H}^{'}_{2}\equiv 0$. 
The total Hamiltonian $\bar{H}=H_{1}+\tilde{H}_{2}$ can
be simplified as
\begin{eqnarray}
\bar{H} &=& \displaystyle{-i\hbar v_{F}\sum_{\sigma}\int dx
[\bar{\psi}^{\dagger}_{R\sigma}(x)\partial_{x}\bar{\psi}_{R\sigma}(x)
-\bar{\psi}^{\dagger}_{L\sigma}(x)\partial_{x}\bar{\psi}_{L\sigma}(x)]}
\nonumber \\
&+& \displaystyle{ \frac{V_{1}}{2}\sum_{\sigma}\int dx
[\alpha\bar{\rho}_{R\sigma}(x)
\bar{\rho}_{R\sigma}(-x)+\alpha\bar{\rho}_{L\sigma}(x)
\bar{\rho}_{L\sigma}(-x)+2\beta\bar{\rho}_{R\sigma}(x)
\bar{\rho}_{L\sigma}(x)]} \label{12} \\
&+& \displaystyle{ \frac{V_{2}}{2}\sum_{\sigma}\int dx
[\alpha\bar{\rho}_{R\sigma}(x)
\bar{\rho}_{R-\sigma}(-x)+\alpha\bar{\rho}_{L\sigma}(x)
\bar{\rho}_{L-\sigma}(-x)+2\beta\bar{\rho}_{R\sigma}(x)
\bar{\rho}_{L-\sigma}(x)]} 
\nonumber\end{eqnarray}
where $\alpha=\frac{1}{2}[1-\cos(2\delta)]$, and 
$\beta=\frac{1}{2}[1+\cos(2\delta)]$. For $\delta=0$, without the impurity
scattering, we have $\alpha=0$ and $\beta=1$, the Hamiltonian (\ref{12})
becomes the original one (\ref{1}).
In terms of these new electron fields $\bar{\psi}_{iR(L)\sigma}(x)$, the
interaction Hamiltonian $\bar{H}^{c}_{I}$ (last two terms)
becomes a very simple form, and the
right- and left-moving electrons are completely separated at the critical
points $\delta^{c}=\pm\pi/2$ induced by the backward scattering
potential. However, for a general phase shift $\delta$,
the total Hamiltonian (\ref{12}) becomes little complex and non-local. The
non-locality of the interaction terms
is the most prominent character of the backward scattering of the conduction
electrons on the impurity, which strongly influences the
low energy behavior of the system.

\section{A magnetic impurity scattering}

Now we consider a magnetic impurity scattering, the Kondo interaction term is
\begin{equation}
H_{K}=\sum_{i}J^{i}_{0}[s^{i}_{R}(0)+s^{i}_{L}(0)]\cdot S^{i}+
\sum_{i}J^{i}_{2k_{F}}[s^{i}_{RL}(0)+s^{i}_{LR}(0)]\cdot S^{i}
\label{13}\end{equation}
where ${\bf S}$ is the impurity spin operator ($S=1/2$),
$s^{i}_{R(L)}(0)=\frac{1}{2}\psi^{\dagger}_{R(L)\alpha}(0)
\sigma^{i}_{\alpha\beta}\psi_{R(L)\beta}(0)$, 
$s^{i}_{RL}(0)=\frac{1}{2}\psi^{\dagger}_{R\alpha}(0)
\sigma^{i}_{\alpha\beta}\psi_{L\beta}(0)$, and 
$s^{i}_{LR}(0)=\frac{1}{2}\psi^{\dagger}_{L\alpha}(0)
\sigma^{i}_{\alpha\beta}\psi_{R\beta}(0)$. In terms of the fields 
$\psi_{1(2)\sigma}(x)$, it can be written as
\begin{equation}
H_{K}=\sum_{i}J^{i}_{0}[s^{i}_{1}(0)+s^{i}_{2}(0)]\cdot S^{i}+
\sum_{i}J^{i}_{2k_{F}}[s^{i}_{1}(0)-s^{i}_{2}(0)]\cdot S^{i}
\label{14}\end{equation}
where $s^{i}_{1(2)}(0)=\frac{1}{2}\psi^{\dagger}_{1(2)\alpha}(0)
\sigma^{i}_{\alpha\beta}\psi_{1(2)\beta}(0)$.
Just as for the non-magnetic impurity case, we replacing the interaction
potentials $J^{z}_{0}$ and $J^{z}_{2k_{F}}$ by the phase shifts 
$\bar{\delta}$ and and $\delta$, respectively, where $\bar{\delta}=
\arctan[J^{z}_{0}/(4\hbar v_{s})]$, and $\delta=
\arctan[J^{z}_{2k_{F}}/(4\hbar v_{F})]$. Performing the unitary transformation
\begin{equation}
U'=\exp\{i\frac{2g_{s}\bar{\delta}}{\pi}\Phi_{+s}(0)S^{z}+
i\frac{2\delta}{\pi}\Phi_{-s}(0)S^{z}\}
\label{15}\end{equation}
where $\Phi_{\pm s}(0)=\frac{1}{2}\{
[\Phi_{1\uparrow}(0)-\Phi_{1\downarrow}(0)]\pm
[\Phi_{2\uparrow}(0)-\Phi_{2\downarrow}(0)]$, and taking the gauge 
transformations
\begin{equation}
\psi_{1\sigma}(x)=\bar{\psi}_{1\sigma}(x)e^{i\sigma\theta_{1}}, \;\;\;
\psi_{2\sigma}(x)=\bar{\psi}_{2\sigma}(x)e^{i\sigma\theta_{2}}, \;\;\;
\theta_{1}-\theta_{2}=2\delta S^{z}
\label{16}\end{equation}
where $\sigma=+1$ for spin-up $\uparrow$, and $\sigma=-1$ for spin-down
$\downarrow$, we have the relations
$U'^{\dagger}(H+H_{K})U'=H_{1}+\tilde{H}_{2}+\tilde{H}^{''}_{2}+\bar{H}_{K}$,
where the Hamiltonians $H_{1}$ and $\tilde{H}_{2}$ are the same as that for
the non-magnetic impurity case (\ref{5}) and (\ref{10}), and 
\begin{eqnarray}
\bar{H}_{K} &=& \displaystyle{
\frac{J_{1}D}{2\pi\hbar v_{F}}\{ e^{-i(1+\frac{2g_{s}\bar{\delta}}{\pi})
\Phi_{+s}(0)}e^{-i(1+\frac{2\delta}{\pi})\Phi_{-s}(0)}S^{+}+h.c.\}}
\nonumber \\
&+& \displaystyle{
\frac{J_{2}D}{2\pi\hbar v_{F}}\{ e^{-i(1+\frac{2g_{s}\bar{\delta}}{\pi})
\Phi_{+s}(0)}e^{i(1-\frac{2\delta}{\pi})\Phi_{-s}(0)}S^{+}+h.c.\}}
\label{17} \\
\tilde{H}^{''}_{2} &=& 
\displaystyle{\pm i\frac{V_{1}\sin(2\delta)}{2}S^{z}
\sum_{\sigma}\int^{\infty}_{0}dx
[\bar{\psi}^{\dagger}_{L\sigma}(-x)\bar{\psi}_{R\sigma}(x)-h.c.]
[\bar{\rho}_{R\sigma}(-x)-\bar{\rho}_{L\sigma}(x)]} 
\nonumber \\
&\pm& \displaystyle{i\frac{V_{2}\sin(2\delta)}{2}S^{z}
\sum_{\sigma}\int^{\infty}_{0}dx
[\bar{\psi}^{\dagger}_{L\sigma}(-x)\bar{\psi}_{R\sigma}(x)-h.c.]
[\bar{\rho}_{R-\sigma}(-x)-\bar{\rho}_{L-\sigma}(x)]} 
\nonumber\end{eqnarray}
where $J_{1}+J_{2}=J^{x}_{0}=J^{y}_{0}$, and
$J_{1}-J_{2}=J^{x}_{2k_{F}}=J^{y}_{2k_{F}}$. The Hamiltonian 
$\tilde{H}^{''}_{2}$, similar to $\bar{H}'_{2}$, 
has a conformal dimension $\Delta'>1$, it only 
contributes higher order correction in the calculation correlation functions, 
it can be neglected as first order approximation. In the following discussion
about the magnetic impurity scattering, we only consider the regions around
the fixed points $\delta=0$ and $\delta^{c}=\pm\pi/2$, at these fixed points
the Hamiltonian $\tilde{H}^{``}_{2}$ is zero, it is reasonable to neglect
the Hamiltonian $\tilde{H}^{``}_{2}$ as first order approximation.
Therefore, except the Hamiltonian $\bar{H}_{K}$ for the magnetic impurity
scattering, both for the magnetic and non-magnetic impurity scattering, there
exists the same total bulk Hamiltonian $H_{1}+\tilde{H}_{2}$. It can be easily
understood because for the magnetic impurity scattering the 
$J^{z}_{0}$- and $J^{z}_{2k_{F}}$-term in fact are usual forward and
backward scattering potential terms, respectively. 
The low energy behavior of the system 
is completely determined by the total bulk Hamiltonian $H_{1}+\tilde{H}_{2}$.
The effect of the impurity scattering on the electrons is reflected on
the change of the interactions among the electrons.

\section{ Calculation of electron Green function}

We first use the Hamiltonian  
$\bar{H}$ (\ref{12}) to calculate the Green function of the electron
fields $\bar{\psi}_{iR(L)\sigma}(x,\tau)$. The action of the system can be
written as
\begin{eqnarray}
S &=& \displaystyle{\sum_{\sigma}\int^{1/(k_{B}T)}_{0}d\tau\int dx
\{\bar{\psi}^{\dagger}_{R\sigma}(x,\tau)(\partial_{\tau}-i\partial_{x})
\bar{\psi}_{R\sigma}(x,\tau)} \nonumber \\
&+& \displaystyle{ 
\bar{\psi}^{\dagger}_{L\sigma}(x,\tau)(\partial_{\tau}+i\partial_{x})
\bar{\psi}_{L\sigma}(x,\tau)\}} \nonumber \\
&-& i\displaystyle{\sum_{\sigma}\int^{1/(k_{B}T)}_{0}d\tau\int dx
\{\phi_{R\sigma}(x,\tau)[\bar{\rho}_{R\sigma}(x,\tau)-
\bar{\psi}^{\dagger}_{R\sigma}(x,\tau)\bar{\psi}_{R\sigma}(x,\tau)]}
\nonumber \\
&+& \displaystyle{\phi_{L\sigma}(x,\tau)[\bar{\rho}_{L\sigma}(x,\tau)-
\bar{\psi}^{\dagger}_{L\sigma}(x,\tau)\bar{\psi}_{L\sigma}(x,\tau)]\}}
\label{18} \\
&+& \displaystyle{ 
\frac{V_{1}}{2}\sum_{\sigma}\int^{1/(k_{B}T)}_{0}d\tau \int dx
\{\alpha\bar{\rho}_{R\sigma}(x,\tau)
\bar{\rho}_{R\sigma}(-x,\tau)} \nonumber \\
&+& \displaystyle{\alpha\bar{\rho}_{L\sigma}(x,\tau)
\bar{\rho}_{L\sigma}(-x,\tau)+2\beta\bar{\rho}_{R\sigma}(x,\tau)
\bar{\rho}_{L\sigma}(x,\tau)\}} \nonumber \\
&+& \displaystyle{ 
\frac{V_{2}}{2}\sum_{\sigma}\int^{1/(k_{B}T)}_{0}d\tau \int dx
\{\alpha\bar{\rho}_{R\sigma}(x,\tau)
\bar{\rho}_{R-\sigma}(-x,\tau)} \nonumber \\
&+& \displaystyle{\alpha\bar{\rho}_{L\sigma}(x,\tau)
\bar{\rho}_{L-\sigma}(-x,\tau)+2\beta\bar{\rho}_{R\sigma}(x,\tau)
\bar{\rho}_{L-\sigma}(x,\tau)\}} 
\nonumber\end{eqnarray}
where $T$ is temperature, and
the auxiliary fields $\phi_{R(L)\sigma}(x,\tau)$ are the Lagrangians 
that introduce the constraint conditions $\bar{\rho}_{R(L)\sigma}(x,\tau)=
\bar{\psi}^{\dagger}_{R(L)\sigma}(x,\tau)\bar{\psi}_{R(L)\sigma}(x,\tau)$.
We have chosen $\hbar=v_{F}=1$. 
The Green function of the electrons $\bar{\psi}_{R(L)\sigma}(x,\tau)$
can be written as
\begin{eqnarray}
G_{R(L)\sigma}(x,\tau;x',\tau') &=& \displaystyle{
\frac{1}{\Omega}\int\prod D\bar{\psi}
\prod D\bar{\psi}^{\dagger}\prod D\bar{\rho}\prod D\phi
\bar{\psi}^{\dagger}_{R(L)\sigma}(x',\tau')\bar{\psi}_{R(L)\sigma}(x,\tau)
e^{-S}} \nonumber \\
&=& \displaystyle{
\frac{1}{\Omega}\int\prod D\bar{\psi}
\prod D\bar{\psi}^{\dagger}\prod D\bar{\rho}\prod D\phi
G_{R(L)\sigma}(x,\tau;x',\tau',[\phi]) e^{-S}}
\nonumber\\
&=& \displaystyle{
\frac{1}{\Omega}\int\prod D\bar{\psi}
\prod D\bar{\psi}^{\dagger}\prod D\bar{\rho}\prod D\phi
G_{R(L)\sigma}(x,\tau;x',\tau',[\rho]) e^{-S}}
\label{19}\end{eqnarray}
where $\Omega=\int\prod D\bar{\psi}\prod D\bar{\psi}^{\dagger}\prod D\bar{\rho}
\prod D\phi e^{-S}$ is the partition functional of the system,
$\prod DX=\prod_{\sigma}DX_{R\sigma}(x,\tau)DX_{L\sigma}(x,\tau)$, where
$X=(\bar{\psi}^{\dagger},\bar{\psi},\bar{\rho},\phi)$. 
The Green function $G_{R(L)\sigma}(x,\tau;x',\tau',[\rho])$ can be obtained
by using the boson representation of the electron fields 
$\bar{\psi}_{R(L)\sigma}(x,\tau)$.
The Green functions $G_{R(L)\sigma}(x,\tau;x',\tau',[\phi])$ satisfy
the equation
\begin{equation}
[\partial_{\tau}\mp i\partial_{x}+i\phi_{R(L)\sigma}(x,\tau)]
G_{R(L)\sigma}(x,\tau;x',\tau',[\phi])=-\delta(x-x')\delta(\tau-\tau')
\label{20}\end{equation}
Because this is the first order linear differential equation, we can take the
following Factorization Ansatz\cite{29}
\begin{eqnarray}
G_{R(L)\sigma}(x,\tau;x',\tau',[\phi]) &=& 
G^{0}_{R(L)\sigma}(x-x',\tau-\tau') \nonumber \\
&\cdot& 
\exp\{f_{R(L)\sigma}(x,\tau,[\phi])-f_{R(L)\sigma}(x',\tau',[\phi])\}
\label{191}\end{eqnarray}
where $[\partial_{\tau}\mp i\partial_{x}]
G^{0}_{R(L)\sigma}(x,\tau)=-\delta(x)\delta(\tau)$, and the fields
$f_{R(L)\sigma}(x,\tau,[\phi])$ satisfy the equation
\begin{equation}
[\partial_{\tau}\mp i\partial_{x}]f_{R(L)\sigma}(x,\tau,[\phi])
=-i\phi_{R(L)\sigma}(x,\tau)
\end{equation}
which can be easily solved\cite{30}
$f_{R(L)\sigma}(x,\tau,[\phi])=k_{B}T\sum_{n}\int \frac{dp}{2\pi}
f_{R(L)\sigma}(p,\omega_{n},[\phi])e^{i(px-\omega_{n}\tau)}$, where
$f_{R(L)\sigma}(p,\omega_{n},[\phi])
=i\phi_{R(L)\sigma}(p,\omega_{n})/(i\omega_{n}\mp p)$.
Therefore, in order to calculate the Green function $
G_{R(L)\sigma}(x,\tau;x',\tau')$, we need to know the effective action
$S_{eff.}[\bar{\rho},\phi]$. After introducing the auxiliary fields
$\phi_{R(L)\sigma}(x,\tau)$, there is only the quadratic form of the electron
fields $\bar{\psi}_{R(L)\sigma}(x,\tau)$ in Eq. (\ref{18}), we can integrate
out them and obtain the potential function $W(x,\tau)=Tr\ln(\partial_{\tau}-i
\partial_{x}+i\phi_{R\sigma})+Tr\ln(\partial_{\tau}+i\partial_{x}
+i\phi_{L\sigma})$, which can be calculated by
using the Green functions (\ref{191}) $W(x,\tau)=\frac{i}{2}[\phi_{R\sigma}
(x,\tau)G_{R\sigma}(x,\tau;x'\rightarrow x, \tau'\rightarrow \tau)+
\phi_{L\sigma}(x,\tau)G_{L\sigma}(x,\tau;x'\rightarrow x, 
\tau'\rightarrow\tau)]$. 
After integrating out the fields 
$\psi_{R(L)\sigma}(x,\tau)$, we can obtain the effective action 
\begin{eqnarray}
S_{eff.}[\bar{\rho},\phi] &=& \displaystyle{
k_{B}T\sum_{\sigma}\sum_{n}\int
\frac{dp}{2\pi}[A_{R}|\phi_{R\sigma}(p,\omega_{n})|^{2}+
A_{L}|\phi_{L\sigma}(p,\omega_{n})|^{2}]} \nonumber \\
&-& i\displaystyle{\sum_{\sigma}\int^{1/(k_{B}T)}_{0}d\tau\int dx
\{\phi_{R\sigma}(x,\tau)\bar{\rho}_{R\sigma}(x,\tau)
+ \phi_{L\sigma}(x,\tau)\bar{\rho}_{L\sigma}(x,\tau)\}}
\label{21} \\
&+& \displaystyle{ 
\frac{V_{1}}{2}\sum_{\sigma}\int^{1/(k_{B}T)}_{0}d\tau \int dx
\{\alpha\bar{\rho}_{R\sigma}(x,\tau)
\bar{\rho}_{R\sigma}(-x,\tau)} \nonumber \\
&+& \displaystyle{\alpha\bar{\rho}_{L\sigma}(x,\tau)
\bar{\rho}_{L\sigma}(-x,\tau)+2\beta\bar{\rho}_{R\sigma}(x,\tau)
\bar{\rho}_{L\sigma}(x,\tau)\}} \nonumber \\
&+& \displaystyle{ 
\frac{V_{2}}{2}\sum_{\sigma}\int^{1/(k_{B}T)}_{0}d\tau \int dx
\{\alpha\bar{\rho}_{R\sigma}(x,\tau)
\bar{\rho}_{R-\sigma}(-x,\tau)} \nonumber \\
&+& \displaystyle{\alpha\bar{\rho}_{L\sigma}(x,\tau)
\bar{\rho}_{L-\sigma}(-x,\tau)+2\beta\bar{\rho}_{R\sigma}(x,\tau)
\bar{\rho}_{L-\sigma}(x,\tau)\}} 
\nonumber\end{eqnarray}
where $A_{R(L)}=\mp\frac{1}{4\pi}\frac{p}{i\omega_{n}\mp p}$.
Integrating out the auxiliary fields $\phi_{R(L)\sigma}(x,\tau)$, we can 
obtain the spin and charge collective excitation spectrums
\begin{eqnarray}
\epsilon_{c} &=& \displaystyle{
\pm[1-(\alpha\pm\beta)^{2}\gamma^{2}_{c}]^{1/2}p} \nonumber \\ 
\epsilon_{s} &=& \displaystyle{
\pm[1-(\alpha\pm\beta)^{2}\gamma^{2}_{s}]^{1/2}p} 
\end{eqnarray} 
It is noting that the charge and spin spectrums are influenced by the 
impurity scattering, which induces the exponents of the electron Green 
function and other correlation functions depending on the phase shift 
$\delta$.
After integrating out the fields $\bar{\rho}_{R(L)\sigma}(x,\tau)$ and 
$\phi_{R(L)\sigma}(x,\tau)$ in Eq.(\ref{19}), and
taking the Wick rotation $\tau\rightarrow it, \;\;\tau'\rightarrow it'$, 
we can obtain the electron Green function expression ($\Delta t>0$) 
\begin{eqnarray}
G_{R(L)\sigma}(x,t;x',t') &=& \displaystyle{e^{\pm ik_{F}\Delta x-
Q_{R(L)}(x,t;x',t')}}
\nonumber \\
Q_{R(L)}(x,t;x',t') &=& \displaystyle{
\frac{1}{4}\sum_{j}\ln[(\Delta x\mp\alpha_{j+}\Delta t\pm i\eta)
(\Delta x\mp\alpha_{j-}\Delta t\pm i\eta)]}
\nonumber\\
&+& \displaystyle{
\frac{1}{8}\sum_{j}(\frac{1}{\alpha_{j+}}-1)
\ln[(\Delta x\pm\alpha_{j+}\Delta t\mp i\eta)
(\Delta x\mp\alpha_{j+}\Delta t\pm i\eta)]}
\nonumber\\
&+& \displaystyle{
\frac{1}{8}\sum_{j}(\frac{1}{\alpha_{j-}}-1)
\ln[(\Delta x\pm\alpha_{j-}\Delta t\mp i\eta)
(\Delta x\mp\alpha_{j-}\Delta t\pm i\eta)]}
\nonumber\\
&+& \displaystyle{
\frac{\alpha}{16}\sum_{j}\frac{\gamma_{j}}{\alpha_{j+}}
\ln\left[\frac{(x+x'+\alpha_{j+}\Delta t- i\eta)^{2}
(x+x'-\alpha_{j+}\Delta t+ i\eta)^{2}}{
(2x+i\eta)(2x-i\eta)(2x'+i\eta)(2x'-i\eta)}\right]}
\label{22}\\
&+& \displaystyle{
\frac{\alpha}{16}\sum_{j}\frac{\gamma_{j}}{\alpha_{j-}}
\ln\left[\frac{(x+x'+\alpha_{j-}\Delta t- i\eta)^{2}
(x+x'-\alpha_{j-}\Delta t+ i\eta)^{2}}{
(2x+i\eta)(2x-i\eta)(2x'+i\eta)(2x'-i\eta)}\right]}
\nonumber\\
&+& \displaystyle{
\frac{\alpha\beta^{2}}{32}\sum_{j}\frac{\gamma^{3}_{j}}{\alpha_{j+}}
\ln\left[\frac{(x+x'+\alpha_{j+}\Delta t- i\eta)^{2}
(x+x'-\alpha_{j+}\Delta t+ i\eta)^{2}}{
(2x+i\eta)(2x-i\eta)(2x'+i\eta)(2x'-i\eta)}\right]}
\nonumber\\
&+& \displaystyle{
\frac{\alpha\beta^{2}}{32}\sum_{j}\frac{\gamma^{3}_{j}}{\alpha_{j-}}
\ln\left[\frac{(x+x'+\alpha_{j-}\Delta t- i\eta)^{2}
(x+x'-\alpha_{j-}\Delta t+ i\eta)^{2}}{
(2x+i\eta)(2x-i\eta)(2x'+i\eta)(2x'-i\eta)}\right]}
\nonumber\end{eqnarray}
where, 
$\alpha^{2}_{j\pm}=1-(\alpha\pm\beta)^{2}\gamma^{2}_{j}, j=c,s$,
$\Delta x=x-x'$, $\Delta t=t-t'$,
$\eta$ is the ultraviolet cut-off factor which is proportional to $D^{-1}$. 
It is easily to demonstrate that at the strong coupling critical points 
$\delta^{c}=\pm\pi/2$, $G_{R(L)\sigma}(0,t;0,0)\sim 
t^{-(1/g_{c}+1/g_{s})/2}$, and at $\delta=0$,
$G_{R(L)\sigma}(0,t;0,0)\sim  t^{-(g_{c}+1/g_{c}+g_{s}+1/g_{s})/4}$.
The usual duality relation of the
correlation exponents near the impurity site $x=0$ between the ultraviolet
($\delta=0$) and the infrared fixed points ($\delta^{c}=\pm\pi/2$) is 
also valid even including the spin degrees of the electrons\cite{schmit}. 
However, it is more important that the correlation exponents are depending 
on the phase shift $\delta$, which is consistent with that
both the charge and spin collective excitation spectrums and the interaction 
among electrons are altered by the impurity scattering. 
From Eq. (\ref{22}) we can obtain the density of state of the
electrons which depends on the distance away from the impurity site. At the
strong coupling critical points $\delta^{c}=\pm\pi/2$, we can obtain the 
relation
\begin{equation}
D^{c}_{R(L)}(x,\omega)\sim\left\{ \begin{array}{ll}
\displaystyle{ \omega^{(1/g_{c}+1/g_{s})/2-1}}, & x\rightarrow 0\\
\displaystyle{ \omega^{(g_{c}+1/g_{c}+g_{s}+1/g_{s})/4-1}}, 
& x\rightarrow\infty\end{array}\right.
\label{222}\end{equation}
which is consistent with previous calculations\cite{12,21}. The impurity
backward scattering suppresses the density of state of electrons near the
impurity site $x=0$, but has little influence on the electron density of state
far away from the impurity.

We can also use the Hamiltonians $H_{1}$ (\ref{5}) and $\tilde{H}_{2}$ 
(\ref{10}) to calculate the Green's function of the fermion fields
$\bar{\psi}_{1(2)\sigma}(x)$ by using the same method as above. 
The action of the system can be written as
\begin{eqnarray}
S' &=& \displaystyle{\sum_{\sigma}\int^{1/(k_{B}T)}_{0}d\tau\int dx
\{\bar{\psi}^{\dagger}_{1\sigma}(x,\tau)(\partial_{\tau}-i\partial_{x})
\bar{\psi}_{1\sigma}(x,\tau)} \nonumber \\
&+& \displaystyle{ 
\bar{\psi}^{\dagger}_{2\sigma}(x,\tau)(\partial_{\tau}-i\partial_{x})
\bar{\psi}_{2\sigma}(x,\tau)} \nonumber \\
&-& i\displaystyle{ \phi_{1\sigma}(x,\tau)[\bar{\rho}_{1\sigma}(x,\tau)-
\bar{\psi}^{\dagger}_{1\sigma}(x,\tau)\bar{\psi}_{1\sigma}(x,\tau)]}
\nonumber \\
&-& \displaystyle{i\phi_{2\sigma}(x,\tau)[\bar{\rho}_{2\sigma}(x,\tau)-
\bar{\psi}^{\dagger}_{2\sigma}(x,\tau)\bar{\psi}_{2\sigma}(x,\tau)]}
\nonumber \\
&-& i\displaystyle{ \lambda_{1\sigma}(x,\tau)[\Gamma_{1\sigma}(x,\tau)-
\bar{\psi}^{\dagger}_{1\sigma}(x,\tau)\bar{\psi}_{2\sigma}(x,\tau)]}
\label{23} \\
&-& \displaystyle{i\lambda_{2\sigma}(x,\tau)[\Gamma_{2\sigma}(x,\tau)-
\bar{\psi}^{\dagger}_{2\sigma}(x,\tau)\bar{\psi}_{1\sigma}(x,\tau)]}
\nonumber \\
&+& \displaystyle{\frac{V_{1}}{4}
[\bar{\rho}_{1\sigma}(x,\tau)+\bar{\rho}_{2\sigma}(x,\tau)]
[\bar{\rho}_{1\sigma}(-x,\tau)+\bar{\rho}_{2\sigma}(-x,\tau)]} \nonumber \\
&+& \displaystyle{\frac{V_{2}}{4}
[\bar{\rho}_{1\sigma}(x,\tau)+\bar{\rho}_{2\sigma}(x,\tau)]
[\bar{\rho}_{1-\sigma}(-x,\tau)+\bar{\rho}_{2-\sigma}(-x,\tau)]} \nonumber \\
&-& \displaystyle{ \frac{V_{1}\cos(2\delta)}{4}
[\Gamma_{1\sigma}(x,\tau)+\Gamma_{2\sigma}(x,\tau) ]
[\Gamma_{1\sigma}(-x,\tau)+\Gamma_{2\sigma}(-x,\tau) ]} \nonumber \\
&-& \displaystyle{ \frac{V_{2}\cos(2\delta)}{4}
[\Gamma_{1\sigma}(x,\tau)+\Gamma_{2\sigma}(x,\tau) ]
[\Gamma_{1-\sigma}(-x,\tau)+\Gamma_{2-\sigma}(-x,\tau) ]\}} 
\nonumber \end{eqnarray}
where the auxiliary fields $\lambda_{1(2)\sigma}(x,\tau)$ introduce the
constraint conditions $\Gamma_{1(2)\sigma}(x,\tau)=
\bar{\psi}^{\dagger}_{1(2)\sigma}(x,\tau)\bar{\psi}_{2(1)\sigma}(x,\tau)$.
The Green functions of the fermions $\bar{\psi}_{R(L)\sigma}(x,\tau)$ can be
written as
\begin{eqnarray}
G_{11(22)\sigma}(x,\tau;x',\tau') &=& \displaystyle{
\frac{1}{\Omega}\int\prod D\bar{\psi}
\prod D\bar{\psi}^{\dagger}\prod D\bar{\rho}\prod D\phi
\bar{\psi}^{\dagger}_{1(2)\sigma}(x',\tau')\bar{\psi}_{1(2)\sigma}(x,\tau)
e^{-S'}} \nonumber \\
&=& \displaystyle{
\frac{1}{\Omega}\int\prod D\bar{\psi}
\prod D\bar{\psi}^{\dagger}\prod D\bar{\rho}\prod D\phi
G_{11(22)\sigma}(x,\tau;x',\tau',[\phi,\lambda]) e^{-S'}}
\label{24}\end{eqnarray}
where $G_{11(22)\sigma}(x,\tau;x',\tau',[\phi,\lambda])$ satisfy the 
linear differential equation
\begin{equation}
\left( \begin{array}{ll} 
{\cal D}+i\phi_{\sigma}, & i\lambda_{\sigma}\\
i\lambda_{\sigma}, & {\cal D}+i\phi_{\sigma}
\end{array}\right)\left( \begin{array}{ll} 
G_{11}, & G_{12}\\
G_{21}, & G_{22}
\end{array}\right)=-\left( \begin{array}{ll} 
\delta(x-x')\delta(\tau-\tau'), & 0\\
0, & \delta(x-x')\delta(\tau-\tau')
\end{array}\right)
\label{25}\end{equation}
where ${\cal D}=\partial_{\tau}-i\partial_{x}$, $G_{ij}=G_{ij}(x,\tau;x',
\tau',[\phi,\lambda])$, $i,j=1,2$, and
we have taken $\phi_{1\sigma}(x,\tau)=\phi_{2\sigma}(x,\tau)
=\phi_{\sigma}(x,\tau)$, and $\lambda_{1\sigma}(x,\tau)
=\lambda_{2\sigma}(x,\tau)=\lambda_{\sigma}(x,\tau)$, because in the action
$S'$ (\ref{23}) there only appears the terms 
$\bar{\rho}_{1\sigma}+\bar{\rho}_{2\sigma}$
and $\Gamma_{1\sigma}+\Gamma_{2\sigma}$, after integrating out the fields
$\bar{\rho}_{1(2)\sigma}(x,\tau)$ and $\Gamma_{1(2)\sigma}(x,\tau)$
we have the relations $\phi_{1\sigma}(x,\tau)=\phi_{2\sigma}(x,\tau)$
and $\lambda_{1\sigma}(x,\tau)=\lambda_{2\sigma}(x,\tau)$.
Using the factorization Ansatz
\begin{equation}
\left\{ \begin{array}{ll}
G_{11(22)} = & \displaystyle{
\frac{1}{2}G^{0}(x,\tau;x',\tau')[e^{\tilde{f}_{\sigma}(x,\tau)-
\tilde{f}_{\sigma}(x',\tau')}+e^{\tilde{f}_{\sigma}(x',\tau')-
\tilde{f}_{\sigma}(x,\tau)}]e^{f_{\sigma}(x,\tau)-f_{\sigma}(x',\tau')}} \\
G_{12(21)} = & \displaystyle{
\frac{1}{2}G^{0}(x,\tau;x',\tau')[e^{\tilde{f}_{\sigma}(x,\tau)-
\tilde{f}_{\sigma}(x',\tau')}-e^{\tilde{f}_{\sigma}(x',\tau')-
\tilde{f}_{\sigma}(x,\tau)}]e^{f_{\sigma}(x,\tau)-f_{\sigma}(x',\tau')}} 
\end{array}\right.
\label{26}\end{equation}
where ${\cal D}G^{0}(x,\tau;x',\tau')=-\delta(x-x')\delta(\tau-\tau')$. If
the fields $f_{\sigma}(x,\tau)$ and $\tilde{f}_{\sigma}(x,\tau)$ satisfy the 
equations
\begin{equation}
\left\{\begin{array}{ll}
{\cal D}f_{\sigma}(x,\tau)= & -i\phi_{\sigma}(x,\tau) \\
{\cal D}\tilde{f}_{\sigma}(x,\tau)= & -i\lambda_{\sigma}(x,\tau) 
\end{array}\right.
\end{equation}
we can easily prove that the expression (\ref{26}) is the exact solution of 
the Eq.(\ref{25}). After
integrating out the fields $\bar{\psi}_{1(2)\sigma}(x,\tau)$, 
$\bar{\rho}_{1(2)\sigma}(x,\tau)$ and $\Gamma_{1(2)\sigma}(x,\tau)$, we can
obtain the effective action
\begin{eqnarray}
S_{eff.}[\phi,\lambda] &=& \displaystyle{
k_{B}T\sum_{n}\int\frac{dp}{2\pi}\{
A(p,\omega_{n})[|\phi_{c}(p,\omega_{n})|^{2}+
|\phi_{s}(p,\omega_{n})|^{2}]} \nonumber \\
&+& \displaystyle{
A(p,\omega_{n})[|\lambda_{c}(p,\omega_{n})|^{2}+
|\lambda_{s}(p,\omega_{n})|^{2}]} \nonumber \\
&+& \displaystyle{
\frac{1}{V_{1}+V_{2}}\phi_{c}(p,-\omega_{n})\phi_{c}(p,\omega_{n})
+\frac{1}{V_{1}-V_{2}}\phi_{s}(p,-\omega_{n})\phi_{s}(p,\omega_{n})}
\label{27} \\
&-& \displaystyle{
\frac{1}{(V_{1}+V_{2})\cos(2\delta)}\lambda_{c}(p,-\omega_{n})
\lambda_{c}(p,\omega_{n})-
\frac{1}{(V_{1}-V_{2})\cos(2\delta)}\lambda_{s}(p,-\omega_{n})
\lambda_{s}(p,\omega_{n})\}}
\nonumber\end{eqnarray}
where $A(p,\omega_{n})=\frac{1}{2\pi}\frac{p}{p-i\omega_{n}}$,
$\phi_{c(s)}(p,\omega_{n})=\frac{\sqrt{2}}{2}[
\phi_{\uparrow}(p,\omega_{n})\pm\phi_{\downarrow}(p,\omega_{n})]$, and
$\lambda_{c(s)}(p,\omega_{n})=\frac{\sqrt{2}}{2}[
\lambda_{\uparrow}(p,\omega_{n})\pm\lambda_{\downarrow}(p,\omega_{n})]$.
Using the effective action $S_{eff.}[\phi,\lambda]$ (\ref{27}), we
can obtain the relations after taking the Wick rotation ($\Delta t>0$)
\begin{eqnarray}
<e^{f_{\sigma}(x,t)-f_{\sigma}(x',t')}> &=& \displaystyle{
\left[\frac{(\Delta x-\Delta t+i\eta)^{2}}{
(\Delta x-\alpha_{c}\Delta t+i\eta)
(\Delta x-\alpha_{s}\Delta t+i\eta)}\right]^{1/4}} \nonumber \\
&\cdot & \displaystyle{
\prod_{j}\left[
(\Delta x+\alpha_{j}\Delta t-i\eta)
(\Delta x-\alpha_{j}\Delta t+i\eta)\right]^{-\mu_{j}/4}} \label{28} \\
&\cdot & \displaystyle{
\prod_{j}\left[\frac{
(x+x'+\alpha_{j}\Delta t-i\eta)^{2}
(x+x'-\alpha_{j}\Delta t+i\eta)^{2}}
{(2x+i\eta)(2x-i\eta)(2x'+i\eta)(2x'-i\eta)}
\right]^{-\nu_{j}/8}} \nonumber 
\end{eqnarray}
\begin{eqnarray}
<e^{\tilde{f}_{\sigma}(x,t)-\tilde{f}_{\sigma}(x',t')}> &=& \displaystyle{
\left[\frac{(\Delta x-\Delta t+i\eta)^{2}}{
(\Delta x-\bar{\alpha}_{c}\Delta t+i\eta)
(\Delta x-\bar{\alpha}_{s}\Delta t+i\eta)}\right]^{1/4}} \nonumber \\
&\cdot & \displaystyle{
\prod_{j}\left[
(\Delta x+\bar{\alpha}_{j}\Delta t-i\eta)
(\Delta x-\bar{\alpha}_{j}\Delta t+i\eta)
\right]^{-\bar{\mu}_{j}/4}} \label{29} \\
&\cdot & \displaystyle{
\prod_{j}\left[\frac{
(x+x'+\bar{\alpha}_{j}\Delta t-i\eta)^{2}
(x+x'-\bar{\alpha}_{j}\Delta t+i\eta)^{2}}
{(2x+i\eta)(2x-i\eta)(2x'+i\eta)(2x'-i\eta)}
\right]^{-\bar{\nu}_{j}/8}} \nonumber 
\end{eqnarray}
where $\alpha^{2}_{j}=1-\gamma^{2}_{j}$, 
$\bar{\alpha}^{2}_{j}=1-\cos^{2}(2\delta)\gamma^{2}_{j}$, 
$\mu_{j}=\frac{1}{2}(\frac{1}{\alpha_{j}}-1)$, 
$\bar{\mu}_{j}=\frac{1}{2}(\frac{1}{\bar{\alpha}_{j}}-1)$, 
$\nu_{j}=\frac{\gamma_{j}}{2\alpha_{j}}$, and
$\bar{\nu}_{j}=-\frac{\gamma_{j}\cos(2\delta)}{2\bar{\alpha}_{j}}$,
where $j=c,s$. The Green functions of the fermions $\bar{\psi}_{1(2)\sigma}
(x,t)$ can be written as 
\begin{equation}
G_{11(22)}(x,t;x',t')=G^{0}(x,t;x',t')<e^{f_{\sigma}(x,t)-f_{\sigma}(x',t')}>
<e^{\tilde{f}_{\sigma}(x,t)-\tilde{f}_{\sigma}(x',t')}>
\label{30}\end{equation}
In the impurity-free case, $\delta=0$, the Green functions 
$G_{11(22)}(0,t;0,0)$ have the asymptotic form in the long time limit
\begin{equation}
G_{11(22)}(0,t;0,0)\sim t^{-(g_{c}+1/g_{c})/4-(g_{s}+1/g_{s})/4}
\label{31}\end{equation}
At the strong coupling critical points $\delta^{c}=\pm\pi/2$, they have 
the form
\begin{equation}
G^{c}_{11(22)}(0,t;0,0)\sim t^{-\frac{1}{2g_{c}}-\frac{1}{2g_{s}}}
\label{32}\end{equation}
It would be pointed out that at $\delta=0$ and the strong coupling critical 
points $\delta^{c}=\pm\pi/2$, the Green functions $G_{R(L)}(0,t;0,0)$ and 
$G_{11(22)}(0,t;0,0)$ both have the same asymptotic form in the long time 
limit which is different from previous mean field approximation calculation.
>From Eq. (\ref{32}), we can obtain the density of state of the fermions
$\bar{\psi}_{1(2)\sigma}(x)$ at the strong coupling critical points 
$\delta^{c}=\pm\pi/2$
\begin{equation}
D^{c}_{1(2)}(0,\omega)\sim \omega^{(1/g_{c}+1/g_{s})/2-1}
\label{999}\end{equation}
which is consistent with Eq. (\ref{222}) for the $x\rightarrow 0$ case.
It is clearly shown in (\ref{222}) and (\ref{999}) that the density of state 
of electrons near the impurity is suppressed by the backward scattering, but
the density of state of electrons far away from the impurity remains intact.

\section{Scattering of electrons on the impurity site x=0}

We now study the scattering of electrons on the impurity site $x=0$. There is
some controversy on this topics. One usually believes that at zero 
temperature the electrons are completely reflected on the impurity site, 
therefore there exist the boundary conditions $\psi_{R\sigma}(x,t)
=\pm\psi_{L\sigma}(-x,t)$,
because the backward scattering potential is renormalized to infinity as
temperature going to zero. However, our exact solution of the Green functions
$G_{11}(x,t;x',t')$ and $G_{22}(x,t;x',t')$ in (\ref{30}) 
both have the same expressions even at the 
strong coupling critical points $\delta^{c}=\pm\pi/2$. Therefore, at zero
temperature even though the electrons are completely reflected on the 
impurity site, it does not mean that the boundary conditions
$\psi_{R\sigma}(x,t)=\pm\psi_{L\sigma}(-x,t)$ are correct. In this section,
we give a correct boundary condition which heavily depends on the phase shift
$\delta$.

The influence of the impurity scattering on the electron fields 
$\psi_{R(L)\sigma}(x)$ is determined by the unitary transformation $U$, after
simple calculation we can obtain the relations
\begin{eqnarray}
U^{\dagger}\psi_{R\sigma}(x)U &=& \displaystyle{
\frac{1}{2}e^{i\theta_{1}}e^{i\frac{\delta}{2}sgn(x)}
\{[1+e^{-i\delta(1+sgn(x))}]\bar{\psi}_{R\sigma}(x)} \nonumber \\
&+& \displaystyle{
[1-e^{-i\delta(1+sgn(x))}]\bar{\psi}_{L\sigma}(-x)\}} \nonumber \\
U^{\dagger}\psi_{L\sigma}(x)U &=& \displaystyle{
\frac{1}{2}e^{i\theta_{1}}e^{-i\frac{\delta}{2}sgn(x)}
\{[1-e^{i\delta(1-sgn(x))}]\bar{\psi}_{R\sigma}(-x)} \label{33} \\
&+& \displaystyle{
[1+e^{i\delta(1-sgn(x))}]\bar{\psi}_{L\sigma}(x)\}} \nonumber 
\end{eqnarray}
where for simplicity we have taken the gauge parameters $\theta_{1(2)}$
satisfying $\theta_{1}-\theta_{2}=\delta$. For more clearly showing the 
influence of the impurity scattering on electron fields, we consider the 
following two cases. One is for the case of $x>0$, the relations (\ref{33})
can be rewritten as
\begin{eqnarray}
U^{\dagger}\psi_{R\sigma}(x)U &=& \displaystyle{
\frac{1}{2}e^{i\theta_{1}}e^{i\frac{\delta}{2}}
[(1+e^{-i2\delta})\bar{\psi}_{R\sigma}(x)
+(1-e^{-i2\delta})\bar{\psi}_{L\sigma}(-x)]} \nonumber \\
U^{\dagger}\psi_{L\sigma}(x)U &=& \displaystyle{
e^{i\theta_{1}}e^{-i\frac{\delta}{2}}\bar{\psi}_{L\sigma}(x)}  
\label{34}\end{eqnarray}
Another one for the case of $x<0$, 
the relations (\ref{33}) can be rewritten as
\begin{eqnarray}
U^{\dagger}\psi_{R\sigma}(x)U &=& \displaystyle{
e^{i\theta_{1}}e^{-i\frac{\delta}{2}}\bar{\psi}_{R\sigma}(x)}  
\label{35} \\
U^{\dagger}\psi_{L\sigma}(x)U &=& \displaystyle{
\frac{1}{2}e^{i\theta_{1}}e^{i\frac{\delta}{2}}
[(1-e^{i2\delta})\bar{\psi}_{R\sigma}(-x)
+(1+e^{i2\delta})\bar{\psi}_{L\sigma}(x)]} \nonumber 
\end{eqnarray}
The physical explanation of Eqs. (\ref{34}) and (\ref{35}) is that a 
right-moving electron from $-\infty$ to $+\infty$ and a left-moving electron
from $+\infty$ to $-\infty$ are reflected at the impurity site $x=0$. for a
general phase shift $\delta$, the right- and left-moving electrons are only
partially reflected on the impurity. However, at the strong coupling critical
points $\delta^{c}=\pm\pi/2$, the right- and left-moving electrons are
completely reflected on the impurity site. This can be easily shown from 
Eqs. (\ref{34}) and (\ref{35})
\begin{eqnarray}
U^{\dagger}\psi_{R\sigma}(x)U|_{\delta^{c}} &=& 
\left\{ \begin{array}{ll} \displaystyle{
e^{i\theta_{1}}e^{i\frac{\pi}{4}}
\bar{\psi}_{L\sigma}(-x)}, & x>0 \\
\displaystyle{ e^{i\theta_{1}}e^{-i\frac{\pi}{4}}
\bar{\psi}_{R\sigma}(x)}, & x<0 \end{array}\right. \nonumber \\
U^{\dagger}\psi_{L\sigma}(x)U|_{\delta^{c}} &=& 
\left\{ \begin{array}{ll} \displaystyle{
e^{i\theta_{1}}e^{-i\frac{\pi}{4}}
\bar{\psi}_{L\sigma}(x)}, & x>0 \\
\displaystyle{ e^{i\theta_{1}}e^{i\frac{\pi}{4}}
\bar{\psi}_{R\sigma}(-x)}, & x<0 \end{array}\right. 
\label{36}\end{eqnarray}
It can be easily seen that there exists a relative phase shift
$\Delta\delta=\pi/2$ between in and out electron wave functions
as the right- and left-moving electrons are completely
reflected on the impurity site $x=0$, which is different from the boundary
conditions $\psi_{R\sigma}(x,t)=\pm\psi_{L\sigma}(-x,t)$.
Therefore, Eq. (\ref{36}) means that at zero temperature this infinity 
one-dimensional system breaks into two half-infinity subsystems 
at the impurity site $x=0$, but the electron fields have a twisted boundary
condition. Eq. (\ref{36}) is consistent with the calculation
of the Green function of the fields $\bar{\psi}_{1(2)\sigma}(x,t)$ at the
strong coupling critical points $\delta^{c}=\pm\pi/2$.

\section{Fermi-edge singularity function of X-ray absorption}

There is some debating about the Fermi-edge singularity of X-ray absorption
because previous perturbation approximation calculations give different
singularity exponents. Now we re-calculate the exponent of the Fermi-edge
singularity function of X-ray absorption which is determined by the 
correlation function
\begin{eqnarray}
I_{\sigma}(t) &=& \displaystyle{
<e^{iHt}\psi^{\dagger}_{1\sigma}(0)e^{-i(H+H_{im})t}\psi_{1\sigma}(0)>}
\nonumber \\
&=& \displaystyle{
<P(t)\bar{\psi}^{\dagger}_{1\sigma}(0,t)U(t)
U^{\dagger}(0)\bar{\psi}_{1\sigma}(0,0)>}
\label{37}\end{eqnarray}
where $P(t)=e^{iHt}e^{-i(H_{1}+\tilde{H}_{2})t}\sim 1$, and
$U(t)=e^{i(H_{1}+\tilde{H}_{2})t}Ue^{-i(H_{1}+\tilde{H}_{2})t}$. 
In order to calculate the correlation function (\ref{37}), we need to know
the correlation functions of the boson fields $\Phi_{\pm s}(x,t)$ and
$\Phi_{\pm c}(x,t)$ near the impurity site $x=0$, 
where $\Phi_{\pm c}(x,t)=\frac{1}{2}\{
[\Phi_{1\uparrow}(x,t)+\Phi_{1\downarrow}(x,t)]\pm 
[\Phi_{2\uparrow}(x,t)+\Phi_{2\downarrow}(x,t)]\}$. 
According to the Hamiltonians $H_{1}$ and $\tilde{H}_{2}$, the correlation 
functions of the boson fields $\Phi_{+c(s)}(x,t)$ are completely determined
by the Hamiltonian
\begin{equation}
H'=\frac{\hbar v_{c}}{4\pi}\int dx 
[\partial_{x}\bar{\Phi}_{+c}(x)]^{2}+\frac{\hbar v_{s}}{4\pi}\int dx
[\partial_{x}\bar{\Phi}_{+s}(x)]^{2} 
\label{38}\end{equation}
where $\bar{\Phi}_{+c}(x)=\cosh(\chi_{c})\Phi_{+c}(x)-\sinh(\chi_{c})
\Phi_{+c}(-x)$, and $\bar{\Phi}_{+s}(x)=\cosh(\chi_{s})
\Phi_{+s}(x)-\sinh(\chi_{s})\Phi_{+s}(-x)$, where the parameters $\chi_{c(s)}$
are defined as $\tanh(2\chi_{c(s)})=\gamma_{c(s)}$. It is worth noting that
the Hamiltonian (\ref{38}) is independent of the impurity scattering.
By simple calculation, we can obtain the correlation functions in the long 
time limit
\begin{eqnarray}
\displaystyle{<e^{-i\Phi_{+c}(0,t)}e^{i\Phi_{+c}(0,0)}>} &\sim& t^{-1/g_{c}}
\nonumber \\
\displaystyle{<e^{-i\Phi_{+s}(0,t)}e^{i\Phi_{+s}(0,0)}>} &\sim& t^{-1/g_{s}}
\label{39}\end{eqnarray}
Using the Green functions $G_{11(22)}(0,t;0,0)$ and Eq. (\ref{39}), 
we can determine the correlation functions of the boson fields
$\Phi_{-c(s)}(0,t)$, because the fermion fields $\bar{\psi}_{1(2)\sigma}(x)$
can be written as in terms of the boson fields $\Phi_{\pm c}(x)$ and
$\Phi_{\pm s}(x)$
\begin{equation}
\bar{\psi}_{1(2)\sigma}(x)\sim\exp\{ i\frac{1}{2}[\Phi_{+c}(x)\pm\Phi_{-c}(x)
+\sigma\Phi_{+s}(x)\pm\sigma\Phi_{-s}(x)]\}
\nonumber\end{equation}
Comparing the Green functions $G_{11(22)}(0,t;0,0)$ (\ref{30}) with Eq. 
(\ref{39}), we can easily obtain the correlation functions 
in the long time limit
\begin{eqnarray}
\displaystyle{<e^{-i\Phi_{-c}(0,t)}e^{i\Phi_{-c}(0,0)}>} &\sim& 
t^{-\bar{g}_{c}}
\nonumber \\
\displaystyle{<e^{-i\Phi_{-s}(0,t)}e^{i\Phi_{-s}(0,0)}>} &\sim& 
t^{-\bar{g}_{s}}
\label{40}\end{eqnarray}
where $\bar{g}_{c(s)}=\sqrt{1-\cos(2\delta)\gamma_{c(s)}}/
\sqrt{1+\cos(2\delta)\gamma_{c(s)}}$. Using Eqs. (\ref{39}) and (\ref{40}),
we can obtain the Fermi-edge singularity function of X-ray absorption
\begin{equation}
I_{\sigma}(\omega)\sim\omega^{\kappa}, \;\;\; \kappa=-1+
\frac{1}{4}(\frac{1}{g_{c}}+\frac{1}{g_{s}}+(1-\frac{2|\delta|}{\pi})^{2}
\bar{g}_{c}+\bar{g}_{s}).
\label{41}\end{equation}
It is noting that the exponent of the Fermi-edge singularity function 
depends on the phase shift $\delta$, therefore the impurity scattering 
influences its low energy behavior. 
At the strong coupling critical points $\delta^{c}=\pm\pi/2$, the exponent
$\kappa$ takes the value $\kappa^{c}=-1+\frac{1}{4g_{c}}+\frac{1}{2g_{s}}$.
For the interaction-free case, $g_{c}=g_{s}=1$, it takes $\kappa^{c}=-1/4$.
For the spinless repulsive interacting fermion system
($V_{2}=0$, $g_{c}=g_{s}=g$, and $\bar{g}_{c}=\bar{g}_{s}$), using the same 
method as above, at the strong coupling critical points $\delta^{c}=\pm\pi/2$
we can easily obtain 
$\kappa^{c}=-\frac{7}{8}+\frac{1}{2g}$. For a free spinless fermion system,
$g=1$, the $\kappa^{c}$ is $-3/8$. This Fermi-edge singularity should be seen
in future experiment.  
These results at $\delta^{c}=\pm\pi/2$
are the same as that in Ref.\cite{3,4,5,6}, and different from
that in Ref.\cite{7} for the interaction-free case.
In these two cases, for small repulsive interaction of the electrons, 
i.e., $g\sim 1$ (spinless), or $g_{c}\sim g_{s}\sim 1$,
there is the Fermi-edge singularity produced by the
backward scattering of the deep core-level hole. 
However, a stronger repulsive interaction of electrons will sweep off
(i.e., $\kappa\geq 0$) the
Fermi-edge singularity induced by the backward scattering of the deep
core-level hole.

\section{Friedel oscillation and charge neutrality}

This is another important issue of the impurity scattering in one-dimensional
system which shows new character different from that in high dimensional
system, where the impurity can be a point-like testing charge. Now we take
the following impurity scattering Hamiltonian
\begin{eqnarray}
H^{'}_{im} &=& \displaystyle{ U(0)\sum_{\sigma}[\rho_{R\sigma}(0)+
\rho_{L\sigma}(0)]} \nonumber \\
&+& \displaystyle{ U(2k_{F})[\psi^{\dagger}_{R\sigma}(0)\psi_{L\sigma}(0)
+\psi^{\dagger}_{L\sigma}(0)\psi_{R\sigma}(0)]}
\label{42}\end{eqnarray}
where $U(0)=Q_{test}(V_{1}+V_{2})$ is usual forward scattering potential, 
$U(2k_{F})=Q_{test}V_{2k_{F}}$ is the backward scattering potential
and $Q_{test}$ is a test charge residing at $x=0$. Now the unitary
transformation $U$ (\ref{7}) is replaced by the following one
\begin{equation}
\bar{U}=\exp\{i\sum_{\sigma}\{\frac{g_{c}\tilde{\delta}}{2\pi}
[\Phi_{1\sigma}(0)+\Phi_{2\sigma}(0)]+\frac{\delta}{2\pi}
[\Phi_{1\sigma}(0)-\Phi_{2\sigma}(0)]\}\}
\label{43}\end{equation}
where the phase shifts $\tilde{\delta}$ and $\delta$ are defined as
$\tilde{\delta}=\arctan[U(0)/(\hbar v_{c})]$, and
$\delta=\arctan[U(2k_{F})/(\hbar v_{F})]$. It must be reminded that the 
forward scattering does not alter the interaction among electrons, therefore
the Green functions of electrons we obtained remain invariant after including
the forward scattering term. 

The total density field of electrons reads
\begin{eqnarray}
\rho(x) &=& \displaystyle{ \frac{1}{2}\sum_{\sigma}[\rho_{R\sigma}(x)+
\rho_{L\sigma}(x)]} \nonumber \\
&+& \displaystyle{ \frac{1}{2}\sum_{\sigma}[e^{-i2k_{F}x}
\psi^{\dagger}_{R\sigma}(x)\psi_{L\sigma}(x)+e^{i2k_{F}x}
\psi^{\dagger}_{L\sigma}(x)\psi_{R\sigma}(x)]}
\label{44}\end{eqnarray}
In terms of the electron fields $\bar{\psi}_{R(L)\sigma}(x)$, we can easily
obtain the relations which describe the influence of the impurity scattering
on the electron density field
\begin{eqnarray}
\rho_{>}(x) &=& \displaystyle{ \bar{U}^{\dagger}\rho(x\geq 0)\bar{U}}
\nonumber \\
&=& \displaystyle{ -\frac{g_{c}\tilde{\delta}}{\pi}\delta(x)+
\frac{1}{2}\sum_{\sigma}[\beta\bar{\rho}_{R\sigma}(x)+\bar{\rho}_{L\sigma}(x)
+\alpha\bar{\rho}_{L\sigma}(-x)]} \nonumber \\
&+& \displaystyle{ i\frac{\sin(2\delta)}{4}[\bar{\psi}^{\dagger}_{R\sigma}(x)
\bar{\psi}_{L\sigma}(-x)-\bar{\psi}^{\dagger}_{L\sigma}(-x)
\bar{\psi}_{R\sigma}(x)]} \label{45} \\
&+& \displaystyle{ \frac{1}{2}\{ 
[\cos(\delta)\bar{\psi}^{\dagger}_{R\sigma}(x)
\bar{\psi}_{L\sigma}(x)-i\sin(\delta)\bar{\psi}^{\dagger}_{L\sigma}(-x)
\bar{\psi}_{L\sigma}(-x)]e^{-i2k_{F}x-ig_{c}\tilde{\delta}} +h.c.\}} 
\nonumber\end{eqnarray}
\begin{eqnarray}
\rho_{<}(x) &=& \displaystyle{ \bar{U}^{\dagger}\rho(x< 0)\bar{U}}
\nonumber \\
&=& \displaystyle{ 
\frac{1}{2}\sum_{\sigma}[\beta\bar{\rho}_{L\sigma}(x)+\bar{\rho}_{R\sigma}(x)
+\alpha\bar{\rho}_{R\sigma}(-x)]} \nonumber \\
&+& \displaystyle{ i\frac{\sin(2\delta)}{4}[\bar{\psi}^{\dagger}_{L\sigma}(x)
\bar{\psi}_{R\sigma}(-x)-\bar{\psi}^{\dagger}_{R\sigma}(-x)
\bar{\psi}_{L\sigma}(x)]} \label{46} \\
&+& \displaystyle{ \frac{1}{2}\{ 
[\cos(\delta)\bar{\psi}^{\dagger}_{R\sigma}(x)
\bar{\psi}_{L\sigma}(x)+i\sin(\delta)\bar{\psi}^{\dagger}_{R\sigma}(x)
\bar{\psi}_{R\sigma}(-x)]e^{-i2k_{F}x+ig_{c}\tilde{\delta}} +h.c.\}} 
\nonumber\end{eqnarray}
At the strong coupling critical points $\delta^{c}=\pm\pi/2$, using the
Green functions of the electron fields $\bar{\psi}_{R(L)\sigma}(x)$, we can
easily obtain the electron density
\begin{equation}
<\rho^{c}_{>,<}(x)> \sim \frac{\cos(2k_{F}x\pm g_{c}\tilde{\delta})}
{x^{(g_{c}+g_{s})/2}}
\label{47}\end{equation}
which shows the unusual Friedel oscillation different from that in high
dimensional system. The total charge induced by the testing charge $Q_{test}$
can be obtained 
\begin{eqnarray}
Q^{c} &=& \int dx <\bar{U}^{\dagger}\rho(x)\bar{U}> \nonumber \\
&=& \displaystyle{ -\frac{g_{c}\tilde{\delta}}{\pi}-a\int^{\infty}_{0}dx
\frac{\sin(g_{c}\tilde{\delta})\sin(2k_{F}x)+ 
\cos(g_{c}\tilde{\delta})\cos(2k_{F}x)}{x^{(g_{c}+g_{s})/2}}}
\label{48}\end{eqnarray}
where $a$ is a constant, and we have chosen $\sum_{\sigma}\int dx[
\bar{\rho}_{R\sigma}(x)+\bar{\rho}_{L\sigma}(x)]=0$. 
In the case of $\tilde{\delta}\sim 0$, i.e., weak
forward scattering, $Q^{c}$ can be rewritten as
\begin{equation}
Q^{c}=-(1-g^{2}_{c})Q_{test}[1+a\pi\int^{\infty}_{0}dx
\frac{\sin(2k_{F}x)}{x^{(g_{c}+g_{s})/2}}]-
a\int^{\infty}_{0}dx
\frac{\cos(2k_{F}x)}{x^{(g_{c}+g_{s})/2}}
\label{49}\end{equation}
The last term derives from the backward scattering, because we have taken
the phase shift $\delta^{c}=\pm\pi/2$ (corresponding to $Q_{test}U(2k_{F})
\rightarrow \pm\infty$), 
it is independent of the testing 
charge $Q_{test}(\neq 0)$. The charge neutrality means that the relation
is rigorously satisfied, $Q^{c}=Q_{test}$. However, Eq. (\ref{49}) shows that
$Q^{c}$ depends on some parameters such as the bandwidth and the size of the 
system. Therefore, the conduction electrons cannot completely screen the 
test charge. Some boundary charge is needed to retain the charge neutrality
of the system. This conclusion is consistent with the previous 
calculation\cite{egger}.
 
\section{Impurity susceptibility of magnetic impurity scattering}

The magnetic impurity scattering described by the Kondo interaction term, 
which is
different from the non-magnetic impurity scattering, we cannot use the unitary
transformation $U'$ (\ref{15}) completely to eliminate the Kondo interaction
term. The spin-exchange interaction term $\bar{H}_{K}$ (\ref{17}) determines
the low energy behavior of the magnetic impurity. Here we only consider the
low temperature dependence of the impurity susceptibility which is completely 
determined by the two phase shifts $\bar{\delta}$ and $\delta$. 
For simplicity, we use a spinless fermion to represent the magnetic impurity
spin, $S^{-}=f$, $S^{+}=f^{\dagger}$, and $S^{z}=f^{\dagger}f-1/2$, the 
Hamiltonian $\bar{H}_{K}$ (\ref{17}) can be written as
\begin{equation}
\bar{H}_{K} = K_{1}[\Psi^{\dagger}(0)f+f^{\dagger}
\Psi(0)]+
K_{2}[\bar{\Psi}^{\dagger}(0)f+f^{\dagger}
\bar{\Psi}(0)]
\label{50}\end{equation}
where $K_{1(2)}=J_{1(2)}(\frac{D}{2\pi\hbar v_{F}})^{1/2}$, 
$\Psi(0)=
(\frac{D}{2\pi\hbar v_{F}})^{1/2}e^{-i(1+\frac{2g_{s}\bar{\delta}}{\pi})
\Phi_{+s}(0)}e^{-i(1+\frac{2\delta}{\pi})\Phi_{-s}(0)}$ and
$\bar{\Psi}(0)=
(\frac{D}{2\pi\hbar v_{F}})^{1/2}e^{-i(1+\frac{2g_{s}\bar{\delta}}{\pi})
\Phi_{+s}(0)}e^{i(1-\frac{2\delta}{\pi})\Phi_{-s}(0)}$ are anyon fields
which anticommutate with the fermion field $f$.
Now we consider three regions determined by the phase shifts $\bar{\delta}$ 
and $\delta$ under the condition $g_{s(c)}\leq 1$, i.e., for the repulsive 
interacting electron system.

\centerline{{\it A. The case of $\bar{\delta}\sim 0$ and 
$\delta\sim\pm\pi/2$}} 

In these regions, the physical property of the system is completely determined
by the Hamiltonian at the critical fixed points $\bar{\delta}=0$ and
$\delta^{c}=\pm\pi/2$. The Hamiltonian $\bar{H}_{K}$ (\ref{50})
can be rewritten as at these critical fixed points
\begin{equation}
\bar{H}_{K} = K[\chi^{\dagger}(0)f+f^{\dagger}\chi(0)]
\label{51}\end{equation}
where $K=K_{1}$ and $\chi(0)=\Psi(0)$ for $\delta=-\pi/2$, and $K=K_{2}$
and $\chi(0)=\bar{\Psi}(0)$ for $\delta=\pi/2$. We have neglected the term 
with high conformal dimension. According to Eqs. (\ref{39}) and (\ref{51}),
we can obtain the Green function of the impurity fermion $f$
\begin{equation}
<f^{\dagger}(\omega)f(-\omega)>\sim\frac{1}{i\omega-\Sigma(\omega)}, \;\;\;
\Sigma(\omega)\sim |\omega|^{-1+1/g_{s}}
\label{52}\end{equation}
If $g_{s}=1$, i.e., the interaction-free electron system, 
the self-energy of the 
impurity fermion $\Sigma(\omega)$ becomes a constant, the system becomes 
usual one-channel Kondo problem. If $g_{s}<1/2$, i.e., the strong interaction
system, the exponent of the self-energy is larger than one, $-1+1/g_{s}>1$,
in the low energy limit the self-energy only contributes higher order 
correction, therefore it can be neglected, the impurity becomes free. However,
If $1/2\leq g_{s}<1$, the low energy behavior of the impurity is determined 
by the self-energy $\Sigma(\omega)$. Based upon above discussion, we can 
obtain the low temperature dependence of the impurity susceptibility
\begin{equation}
\chi_{im}(T)\sim \left\{ \begin{array}{ll}
const., & g_{s}=1\\
\displaystyle{T^{3-2/g_{s}}}, & 1/2\leq g_{s}<1\\
\displaystyle{T^{-1}}, & g_{s}<1/2
\end{array}\right.
\label{53}\end{equation}
Only in the interaction range $1/2<g_{s}<1$, the exponent of the impurity
susceptibility depends on the interaction among electrons. In the strong 
interaction limit, the impurity becomes free. 

\centerline{{\it B. The case of $\bar{\delta}\sim -\pi/2$ 
and $\delta\sim\pm\pi/2$}}

In these regions, the low energy physical property of the impurity is 
determined by the Hamiltonian $\bar{H}_{K}$ at the critical fixed points
$\bar{\delta}^{c}=-\pi/2$ and $\delta^{c}=\pm\pi/2$
\begin{equation}
\bar{H}_{K} = K[\chi^{\dagger}(0)f+f^{\dagger}\chi(0)]
\label{54}\end{equation}
where $\chi(0)=(\frac{D}{2\pi\hbar v_{F}})^{1/2}
\exp\{i(1-g_{s})\Phi_{+s}(0)\}$ has the conformal dimension $(1-g_{s})^{2}/
(2g_{s})$. For the interaction-free case, $g_{s}=1$, the excitation spectrum
of the impurity fermion opens a gap proportional to $K$ due to the interaction
term (\ref{54}). In the case of $g_{s}<1$, we can obtain
the self-energy of the impurity fermion at these 
critical fixed points
\begin{equation}
\Sigma(\omega)\sim |\omega|^{-1+(1-g_{s})^{2}/g_{s}}
\label{55}\end{equation}
and the low temperature dependence of the impurity susceptibility
\begin{equation}
\chi_{im}(T)\sim \left\{ \begin{array}{ll}
\displaystyle{T^{3-2(1-g_{s})^{2}/g_{s}}}, & \bar{g}\leq g_{s}<1\\
\displaystyle{T^{-1}}, & g_{s}<\bar{g}
\end{array}\right.
\label{56}\end{equation}
where the parameter $\bar{g}$ is determined by the equation $(1-\bar{g})^{2}
=2\bar{g}$. 

\centerline{{\it C. The case of $\bar{\delta}\sim -\pi/2$ and $\delta\sim 0$}}

The low energy behavior of the system in this region is determined by the
Hamiltonian $\bar{H}_{K}$ at the critical fixed point $\bar{\delta}^{c}=
-\pi/2$ and $\delta=0$
\begin{equation}
\bar{H}=K_{1}[\chi^{\dagger}(0)f+f^{\dagger}\chi(0)]+
K_{2}[\bar{\chi}^{\dagger}(0)f+f^{\dagger}\bar{\chi(0)}]
\label{56'}\end{equation}
where $\chi(0)=(\frac{D}{2\pi\hbar v_{F}})^{1/2}\exp\{i(1-g_{s})\Phi_{+s}(0)
+i\Phi_{-s}(0)\}$ and 
$\bar{\chi}(0)=(\frac{D}{2\pi\hbar v_{F}})^{1/2}\exp\{i(1-g_{s})\Phi_{+s}(0)
-i\Phi_{-s}(0)\}$. For the interaction-free case, $g_{s}=1$, we have the 
relation $\chi(0)=\bar{\chi}(0)$. The system becomes usual asymmetric 
two-channel Kondo model if $K_{1}\neq K_{2}$. It is worth noting that
the $\bar{\delta}=-\pi/2$ and $\delta=0$ is not a stable critical fixed point,
because for the repulsive interaction system ($g_{s(c)}<1$) the backward 
scattering $J^{z}_{2k_{F}}$-term has the conformal dimension 
$(g_{c}+g_{s})/2<1$, it is relevant in terminology of renormalization group.
The backward scattering potential $J^{z}_{2k_{F}}$ is renormalized to infinity
in low energy limit,
which corresponds to the phase shift $\delta^{c}=\pm\pi/2$. However, 
here we artificially assume that the backward scattering potential is very 
small at some low energy region so that we can discuss some low energy behavior
of the impurity near the point $\delta=0$.

In the case of $g_{s}=1$, the system becomes usual two-channel Kondo problem.
If $K_{1}=K_{2}$, the low temperature dependence of the impurity susceptibility
is\cite{emery} $\chi_{im}(T)\sim \ln(T)$. 
If $K_{1}\neq K_{2}$, the asymmetry of the
two channels destroies the $\ln(T)$ denpendence of the impurity susceptibility
and makes it show the low temperature behavior of one-channel Kondo 
problem\cite{fabrizio}. In the case of $g_{s}<1$, 
the self-energy of the impurity
fermion consists of two parts, one is contributed directly by the anyon fields
$\chi(0)$ and $\bar{\chi}(0)$, and another one is from their hybridization,
\begin{eqnarray}
\Sigma_{1}(\omega) &\sim& \displaystyle{
(K^{2}_{1}+K^{2}_{2})|\omega|^{(1-g_{s})^{2}/g_{s}+g_{s}-1}} \nonumber \\
\Sigma_{2}(\omega) &\sim& \displaystyle{
K_{1}K_{2}|\omega|^{(1-g_{s})^{2}/g_{s}-g_{s}-1}} 
\label{57}\end{eqnarray}
It can be seen that in the range $1/4<g_{s}<1$, the impurity susceptibility
has the power law temperature dependence in the low temperature region. 
However, in the strong interaction region $g_{s}<1/4$, the impurity spin
becomes free, the impurity susceptibility shows the low temperature behavior
$1/T$. It is worth noting that in the three different regions $A,B$ and $C$, 
they all
show that in the strong repulsive interaction region the impurity spin 
becomes free, which is the most prominent character of magnetic impurity
scattering in one-dimensional electron system. These results are consistent 
with previous calculation for a magnetic impurity scattering in Heisenberg 
chain\cite{liu}. The physical explanation of the low temperature power-law
behavior of the impurity susceptibility is that the impurity scattering 
suppresses the density of state of electrons near the impurity site for
the repulsive interacting electron system, therefore the impurity spin is
only partially screened, which is similarly to the non-magnetic impurity
scattering case where the testing charge $Q_{test}$ is partially screened by
conduction electrons.

The impurity susceptibilities in (61) and (64) are different from that in 
Ref.[11]. The result in Ref.[11] was obtained by taking the way that for
$J_{2}=0$ (here we use present labels) and $V_{1}=V_{2}=0$ the system becomes
the usual one-channel Kondo problem, it is well-known that it has 
an infrared Fermi liquid fixed point, then they assumed that after switching
electron interaction, as $\{V_{1},V_{2}\}\rightarrow 0$ the system has the same
infrared fixed point as the one-channel Kondo problem. This assumption is 
crucial, and its correctness is unclear. First, for an one-dimentional 
interacting electron system, the forward scattering potential is not generally 
equal to the backward scattering one because they satisfy different 
renormalization group equations in the low energy limit. Second, for an 
interaction-free electron system, the backward scattering term induced by 
the magnetic impurity is marginal, but for a repulsive interaction electron
system, the backward scattering term is relevant, the 
interaction-free system has an infrared fixed point different from that of the
interaction system.  Generally, there does not exist a principle to guarantee
that there is a smooth connection between these two infrared fixed points as
the interaction potentials $V_{1}$ and $V_{2}$ go to zero.

\section{Conclusion and discussion}

Combining the basic path integral and bosonization methods, we have studied
the low energy behavior of the magnetic and non-magnetic impurity scattering
in an one-dimensional repulsive interacting electron system 
(Tomonaga-Luttinger liquid), and discussed some basic and controversial issues
in this topics. Due to the linearization of the excitation spectrum of 
electrons near their two Fermi levels, using the factorization Ansatz we have 
exactly calculated the Green functions of electrons 
$\bar{\psi}_{R(L)\sigma}(x)$ and fermions $\bar{\psi}_{1(2)\sigma}(x)$ for
a general phase shift $\delta$ induced by the backward scattering potential
of the impurity. The influence of the backward scattering on 
the system is great, because the backward scattering alters the interaction 
among electrons and makes the exponents of all but density-density correlation 
functions depend on the phase shift $\delta$. This the most prominent character
of the backward scattering in Tomonaga-Luttinger liquid. 

Due to the backward
scattering term is relevant, any perturbation expansion method is hard to
give a rigorous description for the system from weak to strong backward
scattering, therefore it is not surprised that there are some controversial
issues in this topics, such as the density of state of electrons near and far
away from the impurity site, the exponent of the Fermi-edge singularity
function of X-ray absorption, the low temperature behavior of the impurity
susceptibility, the boundary conditions of electron fields at zero temperature
on impurity site, and so on. However, by using the simple unitary
and global gauge transformations, the backward scattering term can be 
rigorously
treated, and its influence on the system is incorporated to the interaction
terms among electrons, then using path integral method 
all correlation functions we needed can be exactly calculated. Therefore, we
believe that our results are correct, and can be used to justify previous
results obtained by other methods.

The most important properties of the impurity scattering in Tomonaga-Luttinger
liquid are that: a). At zero temperature the electrons are completely 
reflected on the impurity site $x=0$, the system breaks into two subsystems 
at $x=0$ but the right- and left-moving electron fields have the twisted
boundary condition. b). The density of state
of electrons is suppressed near the impurity site, but it mainly remains
invariance as far away from the impurity. c). The exponents of correlation
functions, such as the Green functions of electrons 
$\bar{\psi}_{R(L)\sigma}(x)$ and fermions $\bar{\psi}_{1(2)\sigma}(x)$, depend
upon the phase shift $\delta$ induced by backward scattering.
d). In the low energy limit, the testing charge is only partially screened
by the conduction electrons.
e). In the weak repulsive interaction region, the impurity susceptibility has
the power-law low temperature dependence. In the strong repulsive
interaction region, the impurity spin becomes free, and the impurity
susceptibility has the $1/T$-type low temperature behavior.

\section{acknowledgment}

The author would like to thank Dr. T. K. Ng for helpful discussions, and 
acknowledge support of HKRGC through Grant No. UST6143/97P.

\end{document}